# A Key Conditional Quotient Filter for Nonlinear, non-Gaussian and non-Markovian System

Yuelin Zhao, *student Member, IEEE*, Feng Wu, *Member, IEEE*, and Li Zhu

*Abstract*—This paper proposes a novel and efficient key conditional quotient filter (KCQF) for the estimation of state in the nonlinear system which can be either Gaussian or non-Gaussian, and either Markovian or non-Markovian. The core idea of the proposed KCQF is that only the key measurement conditions, rather than all measurement conditions, should be used to estimate the state. Based on key measurement conditions, the quotient-form analytical integral expressions for the conditional probability density function, mean, and variance of state are derived by using the principle of probability conservation, and are calculated by using the Monte Carlo method, which thereby constructs the KCQF. Two nonlinear numerical examples were given to demonstrate the superior estimation accuracy of KCQF, compared to seven existing filters.

## I. Introduction

Estimating the true state of a system in the presence of random noise is indeed a common problem, particularly prevalent in fields such as control [1, 2], signal processing [3], position estimation [4, 5] and mechanics [6, 7]. Filtering algorithms are among the commonly employed methods to address this issue. Filtering problems can be categorized into linear filtering problems and nonlinear filtering problems, depending on the different models of the system state and the types of noise. A multitude of theories and algorithms for filtering have been developed to date.

The Kalman filter (KF), proposed by Kalman [8], is designed for linear systems driven by Gaussian noise, which can obtain a minimum mean-squared error estimate of the system state [9]. However, most models encountered in the real world are nonlinear. Therefore, to extend the KF to nonlinear systems, the extended Kalman filter (EKF) has gradually been proposed by researchers [10, 11]. The EKF uses the Taylor series expansion to linearize nonlinear functions and then applies KF for estimation, and hence, requires the system noises to follow Gaussian distributions. It performs well in estimating systems with insignificant nonlinear effects, but can encounter significant errors or even divergence when dealing with systems where nonlinear effects are strong. Currently, researchers have proposed various enhanced algorithms based on EKF, such as [12-17], and so on. Nevertheless, the inherent shortcomings of EKF are still difficult to overcome. To avoid the estimation error caused by linearization of nonlinear system, Julier and Uhlman [18] proposed the unscented Kalman filter (UKF) algorithm. The UKF uses a series of discrete points to approximate the posterior probability density function (PDF), offering higher estimation accuracy. However, the UKF also requires that the noises follow Gaussian distributions. Since the proposal of the UKF, a variety of optimization methods for UKF have also emerged, such as [5, 19-22], and so on. The cubature Kalman filter, proposed by Arasaratnam et al.[23] is also a nonlinear filter. This filter uses a third-order spherical-radial rule to approximate the posterior mean and covariance matrix, making it a more suitable nonlinear filter for high-dimensional systems.

The particle filter (PF) [24] originated from the idea of sequential important sampling based on Bayesian sampling estimation. The PF is a Monte Carlo (MC)-based method that uses a set of weighted particles to approximate the posterior PDF of state. The PF is currently a filter that is well-suited for the systems with strong nonlinearity and non-Gaussian noise. However, the PF suffers from the widely-known problem of particle depletion, wherein a significant fraction of particles loses their weights during update. Although there are some methods to alleviate the particle depletion, such as selecting appropriate importance probability density [25-29] and resampling [30-34], the problem of particle depletion cannot be completely solved.

In summary, each filter has its own characteristics, which implies that different application scenarios might necessitate different filters to achieve optimal outcomes. Especially, these filters mostly assume that the state process is a Markovian process; therefore, when dealing with filtering problems where non-Markovian effects are significant, the accuracy will decrease. Constructing an appropriate filter for nonlinear, non-Gaussian, and non-Markovian systems remains an important topic worthy of research. This paper is committed to developing a filter suitable for nonlinear systems by using the conditional probability density function (CPDF) of the system state. We first theoretically observe, without assumptions such as

This work was supported by the National Natural Science Foundation of China (Nos. 12372190, 62388101), Fundamental Research Funds for the Central Universities (Nos. DUT20RC (5)009, DUT20GJ216), Natural Science Foundation of Liaoning Province (No. 2021-MS-119), Intelligent Multi Source Autonomous Navigation Basic Science Center Open Fund (62388101). *(Corresponding* author: Feng Wu).

Yuelin Zhao, Feng Wu, and Li Zhu are with the State Key Laboratory of Structural Analysis, Optimization and CAE Software for Industrial Equipment, School of Mechanic and Aerospace Engineering, Dalian University of Technology, Dalian 116024, China (e-mail: zhaoyl9811@163.com; wufeng_chn@163.com; zhuli@mail.dlut.edu.cn).

Gaussian distributions and Markov processes, that calculating the CPDF of the state when considering all measurement conditions may suffer from numerical computation instability. Numerical instability arises from the PDF of measurement noise, which is included in both the numerator and denominator of the CPDF expression, and tends to approach zero as the iterations proceed. Therefore, when multiple iterative steps are involved, a situation may arise where a small numerator is divided by a small denominator, leading to significant rounding errors. This observation motivated us to come up with a new idea: the system state should be estimated using key measurement conditions instead of all measurement conditions. Centered around this idea, we use the principle of probability conservation to derive quotient-form analytical expressions for the CPDF, mean, and variance of the state based on key measurement conditions. These analytical expressions, based on key conditions, are in integral form and contain only the PDF of **a** finite number of key measurement noises in both the numerator and denominator, which avoids a small numerator divided by a small denominator, thus making numerical computation more stable. We have also defined the reference value based on the correlation coefficient to help extract key measurement conditions. We employ the MC method to numerically solve the integral terms in the expressions, thus developing an effective key conditional quotient filter (KCQF) suitable for nonlinear systems which can be Gaussian/non-Gaussian, and Markovian/non-Markovian.

The remainder of this article is organized as follows: Section II elaborates on the mathematical model of nonlinear systems; Section III provides a detailed introduction to the KCQF proposed in this paper; Section IV presents two nonlinear test cases, comparing the proposed KCQF with seven existing excellent filters to demonstrate its superiority in estimation accuracy; and Section V concludes with a summary and outlook.

## II. MATHEMATICAL MODELS

Consider the problem of estimating the state of a nonlinear system taking the form

$$\begin{cases} \boldsymbol{x}_{k+1} = \boldsymbol{\varphi}_k(\boldsymbol{x}_k, \boldsymbol{w}_k) \\ \boldsymbol{y}_{k+1} = \boldsymbol{\gamma}_{k+1}(\boldsymbol{x}_{k+1}) + \boldsymbol{v}_{k+1} \end{cases}, \quad (1)$$

where $\boldsymbol{x}_k$ is the system state at time $k$, $\boldsymbol{y}_{k+1}$ is the measurement at time $k+1$, $\boldsymbol{\varphi}_k$ is the nonlinear state transition function, $\boldsymbol{\gamma}_{k+1}$ is the nonlinear measurement function, $\boldsymbol{w}_k$ and $\boldsymbol{v}_{k+1}$ represent zero mean process noise and measurement noise. The joint PDFs of $\boldsymbol{w}_{0:k} = (\boldsymbol{w}_0, \boldsymbol{w}_1, \cdots, \boldsymbol{w}_k)$, and $\boldsymbol{v}_{1:k+1} = (\boldsymbol{v}_1, \boldsymbol{v}_2, \cdots, \boldsymbol{v}_{k+1})$ are denoted by $p_{\boldsymbol{w}_{1:k}}(\boldsymbol{w}_{1:k})$ and $p_{\boldsymbol{v}_{1:k+1}}(\boldsymbol{v}_{1:k+1})$, respectively. Here, noise is no longer assumed to be Gaussian white noise. At the initial moment, the initial state $\boldsymbol{x}_0$ is assumed to follow the distribution $p_{\boldsymbol{x}_0}(\boldsymbol{x}_0)$. The current question is, how to estimate $\boldsymbol{x}_{k+1}$ if a measurement matrix $\boldsymbol{y}_{1:k+1} = (\boldsymbol{y}_1, \boldsymbol{y}_2, \cdots, \boldsymbol{y}_{k+1})$ composed of measurement vectors at different $k$ has already been obtained? For the estimation of $\boldsymbol{x}_{k+1}$, the commonly used method is to solve the CPDF $p_{\boldsymbol{x}_{k+1}|\boldsymbol{y}_{1:k+1}}(\boldsymbol{x}_{k+1} | \boldsymbol{y}_{1:k+1})$. According to Bayesian method [24], $p_{\boldsymbol{x}_{k+1}|\boldsymbol{y}_{1:k+1}}(\boldsymbol{x}_{k+1} | \boldsymbol{y}_{1:k+1})$ can be written as:

$$\begin{aligned} & p_{\boldsymbol{x}_{k+1}|\boldsymbol{y}_{1:k+1}}(\boldsymbol{x}_{k+1} | \boldsymbol{y}_{1:k+1}) \\ & = \frac{p_{\boldsymbol{y}_{k+1}|\boldsymbol{x}_{k+1}}(\boldsymbol{y}_{k+1} | \boldsymbol{x}_{k+1}) p_{\boldsymbol{x}_{k+1}|\boldsymbol{y}_{1:k}}(\boldsymbol{x}_{k+1} | \boldsymbol{y}_{1:k})}{p_{\boldsymbol{y}_{k+1}|\boldsymbol{y}_{1:k}}(\boldsymbol{y}_{k+1} | \boldsymbol{y}_{1:k})} . \end{aligned} \quad (2)$$

If $p_{\boldsymbol{x}_{k+1}|\boldsymbol{y}_{1:k+1}}(\boldsymbol{x}_{k+1} | \boldsymbol{y}_{1:k+1})$ is known, then the mean and variance of $\boldsymbol{x}_{k+1}$ can be expressed as

$$\begin{aligned} \hat{\boldsymbol{x}}_{k+1} &= E(\boldsymbol{x}_{k+1} | \boldsymbol{y}_{1:k+1}) \\ &= \int_{-\infty}^{+\infty} p_{\boldsymbol{x}_{k+1}|\boldsymbol{y}_{1:k+1}}(\boldsymbol{x}_{k+1} | \boldsymbol{y}_{1:k+1}) \boldsymbol{x}_{k+1} \mathrm{d}\boldsymbol{x}_{k+1} \end{aligned}, \quad (3)$$

and

$$\begin{aligned} \boldsymbol{P}_{k+1} &= E\left((\boldsymbol{x}_{k+1} - \hat{\boldsymbol{x}}_{k+1})(\boldsymbol{x}_{k+1} - \hat{\boldsymbol{x}}_{k+1})^{\mathrm{T}} | \boldsymbol{y}_{1:k+1}\right) \\ &= \int_{-\infty}^{+\infty} \left[ \begin{array}{l} p_{\boldsymbol{x}_{k+1}|\boldsymbol{y}_{1:k+1}}(\boldsymbol{x}_{k+1} | \boldsymbol{y}_{1:k+1}) \\ \times (\boldsymbol{x}_{k+1} - \hat{\boldsymbol{x}}_{k+1})(\boldsymbol{x}_{k+1} - \hat{\boldsymbol{x}}_{k+1})^{\mathrm{T}} \end{array} \right] \mathrm{d}\boldsymbol{x}_{k+1} \end{aligned}. \quad (4)$$

The $\hat{\boldsymbol{x}}_{k+1}$ can be used as an estimate of $\boldsymbol{x}_{k+1}$ when $\boldsymbol{y}_{1:k+1}$ is known. Calculating (3) and (4) is challenging because it is often difficult to explicitly express the $p_{\boldsymbol{x}_{k+1}|\boldsymbol{y}_{1:k+1}}(\boldsymbol{x}_{k+1} | \boldsymbol{y}_{1:k+1})$. In existing research, for linear problems, the KF is commonly used for estimation. For nonlinear problems, methods such as EKF, UKF, CKF, and PF can be employed for estimation. However, these methods assume that the state $\boldsymbol{x}_{k+1}$ is Markovian, and the computational accuracy will decrease when dealing with non-Markovian situations.

In the following section, we have precisely derived novel expressions for the CPDF $p_{\boldsymbol{x}_{k+1}|\boldsymbol{y}_{1:k+1}}(\boldsymbol{x}_{k+1} | \boldsymbol{y}_{1:k+1})$, $\hat{\boldsymbol{x}}_{k+1}$ and $\boldsymbol{P}_{k+1}$ without assumptions such as Gaussian distributions and Markov processes, and analyzed that calculating these expressions will lead to numerical instability. To overcome the instability, new quotient-form expressions for the estimation of state based on key measurement conditions have derived, proposing the key conditional quotient filter.

## III. KEY CONDITIONAL QUOTIENT FILTER

### A. A novel quotient-form expression for conditional probability density function

This section proposes a new filter named the key conditional quotient filter (KCQF). To derive this filter, we first present another equivalent form of $p_{\boldsymbol{x}_{k+1}|\boldsymbol{y}_{1:k+1}}(\boldsymbol{x}_{k+1} | \boldsymbol{y}_{1:k+1})$ (equivalent to (2)),

$$p_{\boldsymbol{x}_{k+1}|\boldsymbol{y}_{1:k+1}}(\boldsymbol{x}_{k+1} | \boldsymbol{y}_{1:k+1}) = \frac{p_{\boldsymbol{x}_{k+1},\boldsymbol{y}_{1:k+1}}(\boldsymbol{x}_{k+1}, \boldsymbol{y}_{1:k+1})}{p_{\boldsymbol{y}_{1:k+1}}(\boldsymbol{y}_{1:k+1})} . \quad (5)$$

Equation (5) can be directly given using the CPDF formula. Using (5), a new quotient-form expression for the estimation of $\boldsymbol{x}_{k+1}$ can be provided. To derive this expression, we first introduce the principle of probability conservation. The principle of probability conservation was initially proposed for the study of random dynamical systems [35]. Taking a random dynamical system $u = g(\varepsilon,t)$ as an example, a random interval $D(t) := [\underline{u}, \overline{u}]$ at time $t$, within a short enough time interval $[t, t+\Delta t]$, changes to $D(t+\Delta t) := [\underline{u}(t+\Delta t), \overline{u}(t+\Delta t)]$ as the dynamical system $u(t+\Delta t) = g(\varepsilon, t+\Delta t)$ evolves. If, during $[t, t+\Delta t]$, no new random information is added to this system, and no ran**dom information** is lost, then probability is conserved, which can be expressed as:

$$P(\varphi(\varepsilon,t) \in D(t)) = P(\varphi(\varepsilon, t+\Delta t) \in D(t+\Delta t)). \quad (6)$$

The aforementioned principle of probability conservation essentially states that: a nonlinear mapping $g(\varepsilon,t) \to g(\varepsilon, t+\Delta t)$ that is time-dependent and does not contain additional random information is probability-preserving. From this perspective, any mapping (not necessarily time-dependent), as long as it does not obtain or loss new random information, is also probability-preserving. Based on this idea, we can derive the following more general theorem of principle of probability conservation.

**Lemma 1** *(Principle of Probability Conservation). There exist two random real vectors $z$ and $\theta$, where the PDF of $\theta$ is $p_\theta(\theta)$, and $z$ can be expressed in terms of $\theta$ as follows:*

$$z = (z_1, z_2, \cdots, z_n)^T = (h_1(\theta), h_2(\theta), \cdots, h_n(\theta))^T = h(\theta). \quad (7)$$

*where $h(\theta)$ is a function that only depends on $\theta$ and does not include any other random parameters. Then the joint PDF of $z$ and $\theta$ is:*

$$p_{z,\theta}(z,\theta) = p_\theta(\theta)\delta(z-h(\theta)), \quad (8)$$

*where $\delta(z-h(\theta)) = \prod_{i=1}^{n} \delta(z_i - h_i(\theta))$, and $\delta(\cdot)$ is Dirac function.*

**Proof.** Take any two random vectors $a$ and $b$, whose vector lengths are the same as those of $z$ and $\theta$, respectively, and we analyze the probability of $z \le a \cap \theta \le b$. On the one hand, the probability can be represented by $p_{z,\theta}(z,\theta)$ as

$$P(z \le a \cap \theta \le b) = \int_{-\infty}^{a} \int_{-\infty}^{b} p_{z,\theta}(z,\theta) \mathrm{d}\theta \mathrm{d}z = \int_{-\infty}^{b} \left[\int_{-\infty}^{a} p_{z,\theta}(z,\theta) \mathrm{d}z\right] \mathrm{d}\theta. \quad (9)$$

On the other hand, considering that $z = h(\theta)$ is a function that maps $\theta$ to $z$. In this mapping no other random factors are introduced, therefore the probability of $z \le a \cap \theta \le b$ should be the same as the probability of $h(\theta) \le a \cap \theta \le b$, which means probability conservation, so there is:

$$P(z \le a \cap \theta \le b) = \int_{-\infty}^{+\infty} I(h(\theta) \le a) I(\theta \le b) p_\theta(\theta) \mathrm{d}\theta, \quad (10)$$

where, $I(\cdot)$ is the characteristic function. By combining (9) and (10), we have

$$P(z \le a \cap \theta \le b) = \int_{-\infty}^{b} \left[\int_{-\infty}^{a} p_{z,\theta}(z,\theta) \mathrm{d}z\right] \mathrm{d}\theta$$
$$= \int_{-\infty}^{+\infty} I(h(\theta) \le a) I(\theta \le b) p_\theta(\theta) \mathrm{d}\theta. \quad (11)$$

Taking the derivative of $a$ and $b$ on both sides of (11), there is

$$p_{z,\theta}(a,b) = \delta(a - h(b)) p_\theta(b). \quad (12)$$

At this point, $a$ and $b$ can be replaced by any vector, so letting $a = z$ and $b = \theta$ yields (8). ∎

Using Lemma 1, the following theorem can be derived:

**Theorem 2.** *For (1), when $y_{1:k+1}$ is known, $p_{x_{k+1}|y_{1:k+1}}(x_{k+1} | y_{1:k+1})$ can be expressed as:*

$$p_{x_{k+1}|y_{1:k+1}}(x_{k+1} | y_{1:k+1}) = \frac{\int_{-\infty}^{+\infty} \begin{bmatrix} p_{x_0}(x_0) p_{w_{0:k}}(w_{0:k}) \\ \times p_{v_{1:k+1}}(y_{1:k+1} - \gamma_{1:k+1}(x_{1:k+1})) \\ \times \delta(x_{k+1} - \varphi_k(x_k, w_k)) \end{bmatrix} \mathrm{d}x_0 \mathrm{d}w_{0:k}}{\int_{-\infty}^{+\infty} \begin{bmatrix} p_{x_0}(x_0) p_{w_{0:k}}(w_{0:k}) \\ \times p_{v_{1:k+1}}(y_{1:k+1} - \gamma_{1:k+1}(x_{1:k+1})) \end{bmatrix} \mathrm{d}x_0 \mathrm{d}w_{0:k}}, \quad (13)$$

where $\gamma_{1:k+1}(x_{1:k+1}) = (\gamma_1(x_1), \gamma_2(x_2), \cdots, \gamma_{k+1}(x_{k+1}))$.

**Proof.** According to (1):

$$\begin{aligned} x_{k+1} &= \varphi_k(x_k, w_k) = \varphi_k(\varphi_{k-1}(x_{k-1}, w_{k-1}), w_k) \\ &= \cdots = \varphi_k(\varphi_{k-1}(\cdots \varphi_0(x_0, w_0)), w_k) \\ &= \tilde{\varphi}_k(x_0, w_{0:k}) \\ y_{k+1} &= \gamma_{k+1}(x_{k+1}) + v_{k+1} \end{aligned} \quad (14)$$

According to Lemma 1, there is a joint PDF as follows:

$$p(x_{k+1}, y_{1:k+1}, x_0, v_{1:k+1}, w_{0:k}) = \begin{bmatrix} p_{x_0}(x_0) p_{w_{0:k}}(w_{0:k}) p_{v_{1:k+1}}(v_{1:k+1}) \\ \times \delta(x_{k+1} - \tilde{\varphi}_k(x_0, w_{0:k})) \times \prod_{i=1}^{k+1} \delta(y_i - \gamma_i(x_i) - v_i) \end{bmatrix}. \quad (15)$$

Based on (15), the marginal PDF can be calculated as:

$$\begin{aligned} &p_{x_{k+1}, y_{1:k+1}}(x_{k+1}, y_{1:k+1}) \\ &= \int_{-\infty}^{+\infty} \begin{bmatrix} p_{x_0}(x_0) p_{w_{0:k}}(w_{0:k}) p_{v_{1:k+1}}(v_{1:k+1}) \\ \times \delta(x_{k+1} - \tilde{\varphi}_k(x_0, w_{0:k})) \\ \times \prod_{i=1}^{k+1} \delta(y_i - \gamma_i(x_i) - v_i) \end{bmatrix} \mathrm{d}x_0 \mathrm{d}w_{0:k} \mathrm{d}v_{1:k+1} \\ &= \int_{-\infty}^{+\infty} \begin{bmatrix} p_{x_0}(x_0) p_{w_{0:k}}(w_{0:k}) p_{v_{1:k+1}}(y_{1:k+1} - \gamma_{1:k+1}) \\ \times \delta(x_{k+1} - \tilde{\varphi}_k(x_0, w_{0:k})) \end{bmatrix} \mathrm{d}x_0 \mathrm{d}w_{0:k} \end{aligned} \quad (16)$$

and

$$p_{y_{1:k+1}}(y_{1:k+1})$$
$$=\int_{-\infty}^{+\infty}\begin{bmatrix}p_{x_0}(x_0)p_{w_{0:k}}(w_{0:k})\\ \times p_{v_{1:k+1}}(y_{1:k+1}-\gamma_{1:k+1})\\ \times\delta(x_{k+1}-\tilde{\varphi}_k(x_0,w_{0:k}))\end{bmatrix}dx_0 dw_{0:k}dx_{k+1} \quad .(17)$$
$$=\int_{-\infty}^{+\infty}\left[p_{x_0}(x_0)p_{w_{0:k}}(w_{0:k})p_{v_{1:k+1}}(y_{1:k+1}-\gamma_{1:k+1})\right]dx_0 dw_{0:k}$$

Substituting (16) and (17) into (5) yields (13). ∎

Using Theorem 2, the following corollary can be obtained:

**Corollary 3.** For (1), when $y_{1:k+1}$ is known, the mean and variance of $x_{k+1}$ are

$$\hat{x}_{k+1}=\frac{\int_{-\infty}^{+\infty}\begin{bmatrix}p_{x_0}(x_0)p_{w_{0:k}}(w_{0:k})\\ \times\varphi_k(x_k,w_k)p_{v_{1:k+1}}(y_{1:k+1}-\gamma_{1:k+1})\end{bmatrix}dx_0 dw_{0:k}}{\int_{-\infty}^{+\infty}\begin{bmatrix}p_{x_0}(x_0)p_{w_{0:k}}(w_{0:k})\\ \times p_{v_{1:k+1}}(y_{1:k+1}-\gamma_{1:k+1})\end{bmatrix}dx_0 dw_{0:k}}, (18)$$

and

$$P_{k+1}=\frac{\int_{-\infty}^{+\infty}\begin{bmatrix}p_{x_0}(x_0)p_{w_{0:k}}(w_{0:k})\\ \times p_{v_{1:k+1}}(y_{1:k+1}-\gamma_{1:k+1})\\ \times(\varphi_k(x_k,w_k)-\hat{x}_{k+1})\\ \times(\varphi_k(x_k,w_k)-\hat{x}_{k+1})^T\end{bmatrix}dx_0 dw_{0:k}}{\int_{-\infty}^{+\infty}\begin{bmatrix}p_{x_0}(x_0)p_{w_{0:k}}(w_{0:k})\\ \times p_{v_{1:k+1}}(y_{1:k+1}-\gamma_{1:k+1})\end{bmatrix}dx_0 dw_{0:k}}. \quad (19)$$

The quotient-form mean and variance of $x_{k+1}$ can be calculated using (18) and (19). However, it must be pointed out that calculating the mean and variance of $x_{k+1}$ based on (18) and (19) may result in a very small denominator and numerator, leading to numerical calculation failure. To clearly explain this point, using the example where the measurement noise is white noise. At this point, $p_{v_{1:k+1}}(y_{1:k+1}-\gamma_{1:k+1})=\prod_{i=1}^{k+1}p_{v_i}(y_i-\gamma_i(x_i))$, where the PDF $p_{v_i}(y_i-\gamma_i(x_i^{(j)}))$ represents the likelihood of $y_i-\gamma_i(x_i)$ occurring. Due to $v_i$ represents zero mean noise, $p_{v_i}(y_i-\gamma_i(x_i^{(j)}))$ essentially represents the likelihood of $\gamma_i(x_i)$ being close to $y_i$, and the closer $\gamma_i(x_i)$ and $y_i$ are, the greater the likelihood is. $\prod_{i=1}^{k+1}p_{v_i}(y_i-\gamma_i(x_i))$ represents the likelihood that the measurement $y_i$ is also very close to $\gamma_i(x_i)$ at all moments. Obviously, as $k$ increases, this likelihood will gradually decrease, i.e., $\lim_{k\to\infty}\prod_{i=1}^{k+1}p_{v_i}(y_i-\gamma_i(x_i))=0$. Even if the measurement noise is not white noise, $p_{v_{1:k+1}}(y_{1:k+1}-\gamma_{1:k+1})$ still represents the likelihood that the measurement $y_i$ is very close to $\gamma_i(x_i)$ at all times. Therefore, as $k$ increases, this likelihood will gradually decrease, indicating that $p_{v_{1:k+1}}(y_{1:k+1}-\gamma_{1:k+1})$ approaches 0, which will result in serious rounding errors and numerical instability.

To address the numerical instability issue analyzed above, in the following subsection, we propose a new key conditional quotient filter.

*B. Key Conditional Quotient Filter*

The reason for the instability of the model calculation in the previous subsection is essentially due to too many measurement conditions. Note that (18) and (19) are quotient-form analytical expressions for the estimation of $x_{k+1}$ when $y_{1:k+1}=(y_1,y_2,\cdots,y_{k+1})$ is known. As $k$ increases, $p_{v_{1:k+1}}(y_{1:k+1}-\gamma_{1:k+1})$ will decrease, resulting in smaller numerators and denominators in (18) and (19), leading to significant errors in numerical calculations. In fact, it seems that not all measurement conditions are necessary to estimate $x_{k+1}$. For instance, when estimating $x_{k+1}$, although measurements $y_{1:k+1}=(y_1,y_2,\cdots,y_{k+1})$ have been taken, it is not always necessary to utilize all the measurements. Taking the estimation problem of satellite's trajectory as an example, given the state equation of satellite around the Earth, and the state includes four variables: the radius of the satellite trajectory, the rate of change of the radius, the angle of the satellite trajectory, and the rate of change of the angle. Measurements of the satellite's radius and angle of motion are taken daily to obtain $(y_1,y_2,\cdots,y_{30})$. Now, it is necessary to estimate the state $x_{30}$ at the end of the month. Obviously, the state $x_{30}$ at the end of the month is completely different from $x_1$ at the beginning of the month. Moreover, the impact of the initial measurement $y_1$ on $x_{30}$ is obviously not as critical as the impact of the measurement $y_{30}$ on $x_{30}$. Alternatively, when estimating the radius of the satellite trajectory and the rate of change of the radius, the measurement radius is a more important measurement condition compared to the measurement angle. Thus, the fundamental premise of the KCQF is that, despite having obtained a great number of measurements $y_{1:k+1}=(y_1,y_2,\cdots,y_{k+1})$, to accurately estimate $x_{k+1}$, one should focus on selecting the measurement conditions that are key for $x_{k+1}$ and disregard measurements with weaker correlations.

Assuming that within the already obtained measurements, only a portion of the measurements is key for estimating $x_{k+1}$, it naturally follows that an estimation of the state should be based solely on the key measurement conditions. To concretely illustrate this concept, let's denote the key measurements as $z_{k+1}:=y_{1:k+1}A_{k+1}$. Here, $A_{k+1}$ represents the operational operator for extracting the measurement conditions that are key for the estimation of $x_{k+1}$ from the measurement matrix $y_{1:k+1}$. For instance, if $(y_k,y_{k+1})$ is very important for the estimation
4

of $x_{k+1}$, we can set $A_{k+1} = \begin{bmatrix} 0 & \cdots & 1 & 0 \\ 0 & \cdots & 0 & 1 \end{bmatrix}^T$, thus $z_{k+1} = y_{1:k+1} A_{k+1} = (y_k, y_{k+1})$. The measurement error of $z_{k+1}$ is denoted as $\beta_{k+1} := v_{1:k+1} A_{k+1}$.

To estimate the $x_{k+1}$ based on key measurement conditions $z_{k+1}$, it is necessary to know the CPDF $p_{x_{k+1}|z_{k+1}}(x_{k+1}|z_{k+1})$.

Using Lemma 1, we can establish Theorem 4.

**Theorem 4.** *For (1), when $y_{1:k+1}$, $A_{k+1}$ and the PDF $p_{\beta_{k+1}}(\beta_{k+1})$ of $\beta_{k+1}$ is known, and let $\gamma_{1:k+1}(x_{1:k+1}) = (\gamma_1(x_1), \gamma_2(x_2), \cdots, \gamma_{k+1}(x_{k+1}))$, $z_{k+1} = y_{1:k+1} A_{k+1}$, then the CPDF $p_{x_{k+1}|z_{k+1}}(x_{k+1}|z_{k+1})$ can be expressed as:*

$$p_{x_{k+1}|z_{k+1}}(x_{k+1}|z_{k+1})$$

$$= \frac{\int_{-\infty}^{+\infty} \begin{bmatrix} p_{x_0}(x_0) p_{w_{0:k}}(w_{0:k}) \\ \times p_{\beta_{k+1}}(z_{k+1} - \gamma_{1:k+1}(x_{1:k+1}) A_{k+1}) \\ \times \delta(x_{k+1} - \varphi_k(x_k, w_k)) \end{bmatrix} dx_0 dw_{0:k}}{\int_{-\infty}^{+\infty} \begin{bmatrix} p_{x_0}(x_0) p_{w_{0:k}}(w_{0:k}) \\ \times p_{\beta_{k+1}}(z_{k+1} - \gamma_{1:k+1}(x_{1:k+1}) A_{k+1}) \end{bmatrix} dx_0 dw_{0:k}}. \quad (20)$$

**Proof.** According to (1)

$$\begin{aligned} x_{k+1} &= \varphi_k(x_k, w_k) = \tilde{\varphi}_k(x_0, w_{0:k}) \\ z_{k+1} &= \gamma_{1:k+1}(x_{1:k+1}) A_{k+1} + \beta_{k+1} \end{aligned}. \quad (21)$$

According to Lemma 1, there is a joint PDF as follows:

$$\begin{aligned} & p(x_{k+1}, z_{k+1}, x_0, \beta_{k+1}, w_{0:k}) \\ &= p_{x_0}(x_0) p_{w_{0:k}}(w_{0:k}) p_{\beta_{k+1}}(\beta_{k+1}) \\ &\times \delta(x_{k+1} - \tilde{\varphi}_k(x_0, w_{0:k})) \\ &\times \delta(z_{k+1} - \gamma_{1:k+1}(x_{1:k+1}) A_{k+1} - \beta_{k+1}) \end{aligned}. \quad (22)$$

According to (22), the marginal PDF can be calculated as:

$$p_{x_{k+1}, z_{k+1}}(x_{k+1}, z_{k+1})$$

$$= \int_{-\infty}^{+\infty} \begin{bmatrix} p_{x_0}(x_0) p_{w_{0:k}}(w_{0:k}) p_{\beta_{k+1}}(\beta_{k+1}) \\ \times \delta(x_{k+1} - \tilde{\varphi}_k(x_0, w_{0:k})) \\ \times \delta(z_{k+1} - \gamma_{1:k+1}(x_{1:k+1}) A_{k+1} - \beta_{k+1}) \end{bmatrix} dx_0 dw_{0:k} d\beta_{k+1}, \quad (23)$$

$$= \int_{-\infty}^{+\infty} \begin{bmatrix} p_{x_0}(x_0) p_{w_{0:k}}(w_{0:k}) \\ \times p_{\beta_{k+1}}(z_{k+1} - \gamma_{1:k+1}(x_{1:k+1}) A_{k+1}) \\ \times \delta(x_{k+1} - \tilde{\varphi}_k(x_0, w_{0:k})) \end{bmatrix} dx_0 dw_{0:k}$$

and

$$p_{z_{k+1}}(z_{k+1})$$

$$= \int_{-\infty}^{+\infty} \begin{bmatrix} p_{x_0}(x_0) p_{w_{0:k}}(w_{0:k}) \\ \times p_{\beta_{k+1}}(z_{k+1} - \gamma_{1:k+1}(x_{1:k+1}) A_{k+1}) \\ \times \delta(x_{k+1} - \tilde{\varphi}_k(x_0, w_{0:k})) \end{bmatrix} dx_0 dw_{0:k} dx_{k+1}. \quad (24)$$

$$= \int_{-\infty}^{+\infty} \begin{bmatrix} p_{x_0}(x_0) p_{w_{0:k}}(w_{0:k}) \\ \times p_{\beta_{k+1}}(z_{k+1} - \gamma_{1:k+1}(x_{1:k+1}) A_{k+1}) \end{bmatrix} dx_0 dw_{0:k}$$

Substituting (23) and (24) into $p_{x_{k+1}|z_{k+1}}(x_{k+1}|z_{k+1}) = \dfrac{p_{x_{k+1}, z_{k+1}}(x_{k+1}, z_{k+1})}{p_{z_{k+1}}(z_{k+1})}$ yields (20). ∎

Using Theorem 4, the following corollary can be obtained:

**Corollary 5.** *For (1), when $z_{k+1} = y_{1:k+1} A_{k+1}$, $\beta_{k+1} = v_{1:k+1} A_{k+1}$ and $p_{\beta_{k+1}}(\beta_{k+1})$ is known, the mean and variance of $x_{k+1}$ can be expressed as:*

$$\hat{x}_{k+1|z_{k+1}}$$

$$= \int_{-\infty}^{+\infty} p_{x_{k+1}|z_{k+1}}(x_{k+1}|z_{k+1}) x_{k+1} dx_{k+1} \quad (25)$$

$$= \frac{\int_{-\infty}^{+\infty} \begin{bmatrix} p_{x_0}(x_0) p_{w_{0:k}}(w_{0:k}) \varphi_k(x_k, w_k) \\ \times p_{\beta_{k+1}}(z_{k+1} - \gamma_{1:k+1}(x_{1:k+1}) A_{k+1}) \end{bmatrix} dx_0 dw_{0:k}}{\int_{-\infty}^{+\infty} \begin{bmatrix} p_{x_0}(x_0) p_{w_{0:k}}(w_{0:k}) \\ \times p_{\beta_{k+1}}(z_{k+1} - \gamma_{1:k+1}(x_{1:k+1}) A_{k+1}) \end{bmatrix} dx_0 dw_{0:k}}$$

*and*

$$P_{k+1|z_{k+1}}$$

$$= \int_{-\infty}^{+\infty} \begin{bmatrix} p_{x_{k+1}|z_{k+1}}(x_{k+1}|z_{k+1}) \\ \times (x_{k+1} - \hat{x}_{k+1|z_{1+k}})(x_{k+1} - \hat{x}_{k+1|z_{1+k}})^T \end{bmatrix} dx_{k+1}$$

$$= \frac{\int_{-\infty}^{+\infty} \begin{bmatrix} p_{x_0}(x_0) p_{w_{0:k}}(w_{0:k}) \\ \times p_{\beta_{k+1}}(z_{k+1} - \gamma_{1:k+1}(x_{1:k+1}) A_{k+1}) \\ \times (\varphi_k(x_k, w_k) - \hat{x}_{k+1|z_{1+k}}) \\ \times (\varphi_k(x_k, w_k) - \hat{x}_{k+1|z_{1+k}})^T \end{bmatrix} dx_0 dw_{0:k}}{\int_{-\infty}^{+\infty} \begin{bmatrix} p_{x_0}(x_0) p_{w_{0:k}}(w_{0:k}) \\ \times p_{\beta_{k+1}}(z_{k+1} - \gamma_{1:k+1}(x_{1:k+1}) A_{k+1}) \end{bmatrix} dx_0 dw_{0:k}}. \quad (26)$$

Equations (25) and (26) can be used to calculate the mean and variance of $x_{k+1}$ based on the key conditions. The MC method is utilized to solve the high-dimensional integrals present in (25) and (26). Assuming sampling based on distributions $p_{x_0}(x_0)$ and $p_{w_{0:k}}(w_{0:k})$, $N_s$ samples $x_0^{(j)}$ and $w_{0:k}^{(j)}$, can be obtained, where $j = 1, 2, \cdots, N_s$. By iteratively calculating in terms of (1), $N_s$ samples of state $x_{k+1}^{(j)} = \varphi_k(x_k^{(j)}, w_k^{(j)})$ can be obtained. According to the MC method,

$$\int_{-\infty}^{+\infty} \begin{bmatrix} p_{x_0}(\boldsymbol{x}_0) p_{w_{0:k}}(\boldsymbol{w}_{0:k}) \\ \times p_{\beta_{k+1}}(\boldsymbol{z}_{k+1} - \boldsymbol{\gamma}_{1:k+1}(\boldsymbol{x}_{1:k+1})\boldsymbol{A}_{k+1}) \end{bmatrix} \mathrm{d}\boldsymbol{x}_0 \mathrm{d}\boldsymbol{w}_{0:k}$$

$$\approx \frac{1}{N_s} \sum_{j=1}^{N_s} p_{\beta_{k+1}}(\boldsymbol{z}_{k+1} - \boldsymbol{\gamma}_{1:k+1}(\boldsymbol{x}_{1:k+1}^{(j)})\boldsymbol{A}_{k+1}) \quad , \quad (27)$$

$$\int_{-\infty}^{+\infty} \begin{bmatrix} p_{x_0}(\boldsymbol{x}_0) p_{w_{0:k}}(\boldsymbol{w}_{0:k}) \boldsymbol{\varphi}_k(\boldsymbol{x}_k, \boldsymbol{w}_k) \\ \times p_{\beta_{k+1}}(\boldsymbol{z}_{k+1} - \boldsymbol{\gamma}_{1:k+1}(\boldsymbol{x}_{1:k+1})\boldsymbol{A}_{k+1}) \end{bmatrix} \mathrm{d}\boldsymbol{x}_0 \mathrm{d}\boldsymbol{w}_{0:k}$$

$$\approx \frac{1}{N_s} \sum_{j=1}^{N_s} \begin{bmatrix} \boldsymbol{\varphi}_k(\boldsymbol{x}_k^{(j)}, \boldsymbol{w}_k^{(j)}) \\ \times p_{\beta_{k+1}}(\boldsymbol{z}_{k+1} - \boldsymbol{\gamma}_{1:k+1}(\boldsymbol{x}_{1:k+1}^{(j)})\boldsymbol{A}_{k+1}) \end{bmatrix} \quad , \quad (28)$$

and

$$\int_{-\infty}^{+\infty} \begin{bmatrix} p_{x_0}(\boldsymbol{x}_0) p_{w_{0:k}}(\boldsymbol{w}_{0:k}) \\ \times p_{\beta_{k+1}}(\boldsymbol{z}_{k+1} - \boldsymbol{\gamma}_{1:k+1}(\boldsymbol{x}_{1:k+1})\boldsymbol{A}_{k+1}) \\ \times (\boldsymbol{\varphi}_k(\boldsymbol{x}_k, \boldsymbol{w}_k) - \hat{\boldsymbol{x}}_{k+1|z_{1+k}}) \\ \times (\boldsymbol{\varphi}_k(\boldsymbol{x}_k, \boldsymbol{w}_k) - \hat{\boldsymbol{x}}_{k+1|z_{1+k}})^{\mathrm{T}} \end{bmatrix} \mathrm{d}\boldsymbol{x}_0 \mathrm{d}\boldsymbol{w}_{0:k}$$

$$\approx \frac{1}{N_s} \sum_{j=1}^{N_s} \begin{bmatrix} (\boldsymbol{\varphi}_k(\boldsymbol{x}_k^{(j)}, \boldsymbol{w}_k^{(j)}) - \hat{\boldsymbol{x}}_{k+1|z_{1+k}}) \\ \times (\boldsymbol{\varphi}_k(\boldsymbol{x}_k^{(j)}, \boldsymbol{w}_k^{(j)}) - \hat{\boldsymbol{x}}_{k+1|z_{1+k}})^{\mathrm{T}} \\ \times p_{\beta_{k+1}}(\boldsymbol{z}_{k+1} - \boldsymbol{\gamma}_{1:k+1}(\boldsymbol{x}_{1:k+1}^{(j)})\boldsymbol{A}_{k+1}) \end{bmatrix} \quad . \quad (29)$$

Substitute (27)-(29) into (25) and (26) to calculate $\hat{\boldsymbol{x}}_{k+1|z_{k+1}}$ and $\boldsymbol{P}_{k+1|z_{k+1}}$. Due to the law of large numbers [36], the error order of (27)-(29) is $O(N_s^{-0.5})$. Therefore, as the sample size $N_s$ increases, the results of (27)-(29) will gradually converge to the exact integral.

Corollary 5 provides the quotient-form estimation expressions for the state based on key conditions. What distinguishes these expressions from (18) and (19), which are based on all measurement conditions, is that in the estimation expressions for the state based on key conditions, only the PDFs of some key measurements are used. When calculating using (18) and (19), the high-dimensional joint PDF $p_{v_{1:k+1}}(\boldsymbol{y}_{1:k+1} - \boldsymbol{\gamma}_{1:k+1})$ that appears in the denominator will gradually approach 0 as $k$ increases, leading to calculation failure. When calculating using (25) and (26), $p_{\beta_{k+1}}(\boldsymbol{z}_{k+1} - \boldsymbol{\gamma}_{1:k+1}(\boldsymbol{x}_{1:k+1})\boldsymbol{A}_{k+1})$ is merely the joint PDF of the errors in the key measurements. If the number of key conditions does not increase with the increase of $k$, then $p_{\beta_{k+1}}(\boldsymbol{z}_{k+1} - \boldsymbol{\gamma}_{1:k+1}(\boldsymbol{x}_{1:k+1})\boldsymbol{A}_{k+1})$ will not tend towards 0. Therefore, estimating the state based on key conditions can prevent the issue of dividing a small number by another small number, which effectively ensuring computational stability.

At present, the issue of how to extract key measurements still requires further discussion. This paper uses correlation coefficients to assess the reference value of different measurements for the estimation of state $\boldsymbol{x}_{k+1}$. For instance, when considering whether to utilize the $i$-th measurement data in $\boldsymbol{y}_j$ (denoted as $y_{j,i}$) to estimate the $m$-th state in $\boldsymbol{x}_{k+1}$ (denoted as $x_{k+1,m}$), the reference value of $y_{j,i}$ can be defined as

$$r(y_{i,j}, x_{k+1,m}) = \left| \frac{\mathrm{cov}(y_{i,j}, x_{k+1,m})}{\sqrt{\mathrm{cov}(y_{i,j}, y_{i,j})}\sqrt{\mathrm{cov}(x_{k+1,m}, x_{k+1,m})}} \right|, \quad (30)$$

where $\mathrm{cov}(\cdot, \cdot)$ represents the covariance of two random variables, which can also be conveniently calculated through MC integration. When evaluating $\boldsymbol{x}_{k+1}$, we extrapolate $k'$ steps forward from the current $k+1$ steps, and assume that measurements prior to $k'$ has little reference value for evaluating $\boldsymbol{x}_{k+1}$. Then, using (30), we calculate the reference value of the measurement data within the time steps $k'$ to $k+1$, and select the $d$ measurements with the greatest reference value as the key measurements $\boldsymbol{z}_{k+1}$ for evaluating $\boldsymbol{x}_{k+1}$.

After the key measurement conditions $\boldsymbol{z}_{k+1}$ is determined, its error is $\boldsymbol{\beta}_{k+1} := \boldsymbol{v}_{1:k+1}\boldsymbol{A}_{k+1} = (\beta_{k_1}, \cdots, \beta_{k_d})$, where $\beta_{k_1}, \beta_{k_2}, \cdots, \beta_{k_d}$ represents the measurement noise of the key measurement conditions. In fact, $\beta_{k_1}, \beta_{k_2}, \cdots, \beta_{k_d}$ is composed of $d$ random variables extracted from the measurement noise $\boldsymbol{v}_{1:k+1}$. Therefore, $p_{\beta_{k+1}}(\boldsymbol{\beta}_{k+1})$ is the marginal PDF of $p_{v_{1:k+1}}(\boldsymbol{v}_{1:k+1})$, and can be expressed as:

$$p_{\beta_{k+1}}(\boldsymbol{\beta}_{k+1}) = \int_{-\infty}^{+\infty} p_{v_{1:k+1}}(\boldsymbol{v}_{1:k+1}) \mathrm{d}v_{i_1} \mathrm{d}v_{i_2} \cdots \mathrm{d}v_{i_{k+1-d}}, \quad (31)$$

where $v_{i_1}, v_{i_2}, \cdots, v_{i_{k+1-d}}$ represents the measurement noise of the data that have not been selected as key measurement conditions. When the measurement noise $\boldsymbol{v}_{1:k+1}$ is a Gaussian process or white noise, $p_{\beta_{k+1}}(\boldsymbol{\beta}_{k+1})$ can be easily obtained using (31). When the situation of $p_{v_{1:k+1}}(\boldsymbol{v}_{1:k+1})$ is more complex, MC can also be used to approximate the integration of (31).

This new filter proposed in Section III.B is named key conditional quotient filter (KCQF). Considering the estimation of the mean and variance of $\boldsymbol{x}_{k+1}$ when key measurements are known, the pseudocode of the KCQF is shown in Algorithm 1, in which $K$ is the number of time steps.

In the next section, we will further demonstrate the advantages of KCQF through two nonlinear numerical examples.

---

**Algorithm 1** Key Conditional Quotient Filter

**Input:** $p_{x_0}(\boldsymbol{x}_0)$, $p_{w_{0:k}}(\boldsymbol{w}_{0:k})$, $p_{v_{1:k+1}}(\boldsymbol{v}_{1:k+1})$, $\boldsymbol{y}_{0:K}$, $\boldsymbol{\varphi}_k$,

**Output:** $\hat{\boldsymbol{x}}_{k+1|z_{k+1}}$, $\boldsymbol{P}_{k+1|z_{k+1}}$

Sample according to $p_{x_0}(x_0)$ and $p_{w_{0:k}}(w_{0:k})$ to obtain $N_s$ samples $x_0^{(j)}$ and $w_{0:k}^{(j)}$, $j = 1, 2, \cdots, N_s$;

**for** $k = 1:K$

  **for** $j = 1:N_s$

    Calculate $x_{k+1}^{(j)} = \varphi_k(x_k^{(j)}, w_k^{(j)})$ using (1);

  **end**

**end**

**for** $k = 1:K$

  Using (30) to select $d$ key measurements $z_{k+1}$, and obtain $p_{\beta_{k+1}}(\beta_{k+1})$ based on (31);

  Calculate three high-dimensional integrals using (27)-(29);

  Calculate $\hat{x}_{k+1|z_{k+1}}$ and $P_{k+1|z_{k+1}}$ using (25) and (26);

**end**

## IV. NUMERICAL EXAMPLE

### A. Gaussian Markovian nonlinear example

The first example verifies the computational performance of KCQF proposed in Section III through a widely used Gaussian and Markovian nonlinear numerical example. This example or its variants have been extensively studied [37, 38], and the model is

$$\begin{cases} x_{k+1} = \frac{1}{2}x_k + \frac{25x_k}{1+x_k^2} + 8\cos(1.2k) + w_{k+1} \\ y_{k+1} = \frac{x_{k+1}^2}{20} + v_{k+1}, k = 1, 2, \cdots, K \end{cases} \quad (32)$$

The process noise $w_k$ and measurement noise $v_k$ are assumed to be independent zero mean Gaussian random variables, where $w_k \sim N(0, 10)$, and $v_k \sim N(0, 1)$. The initial value is $x_0 \sim N(0, 2)$ and the number of iteration steps is $K = 52$. A MC averaged root mean squared error $E_{rms}(k)$ is considered for evaluating the accuracy of the estimates. The $E_{rms}(k)$ is computed over a set of MC runs.

$$E_{rms}(k) = \sqrt{\frac{1}{M}\sum_{m=1}^{M}(x^m(k) - \hat{x}^m(k))^2}, \quad (33)$$

where $M$ is the number of MC runs, and in this example, $M = 50$. $x^m(k)$ and $\hat{x}^m(k)$ represent the actual and estimated states at the time instant $k$ during the $m$-th MC run. The time averaged error $\overline{E_{rms}}$ can be calculated by

$$\overline{E_{rms}} = \frac{1}{K}\sum_{k=1}^{K}E_{rms}(k). \quad (34)$$

Firstly, to study the impact of the number of key conditions $d$, we tested the computational performance of the KCQF with $d = 1, 2, 3$ and 4, and plotted the $E_{rms}$ in Fig. 1, where the number of samples $N_s = 50$. For the convenience of discussion, we denote the KCQF with $d$ key conditions as KCQF-$d$. From Fig. 1, it is clear that for this nonlinear model (32), the computational results of KCQF-2 and KCQF-3 are better. When $d = 2$ and 3, the $E_{rms}$ is superior to that when $d = 1$, this indicates that considering more key measurements yields a more accurate estimation result than considering only one key measurement. When $d = 2$ and 3, the $E_{rms}$ is superior to that when $d = 4$. This phenomenon corroborates the point made in section III.B: If too many measurement values are considered, the $p_{\beta_{k+1}}(\beta_{k+1})$ will become too small, leading to both the numerator and the denominator in (18) and (19) to become small, thereby increasing the error in numerical computation.

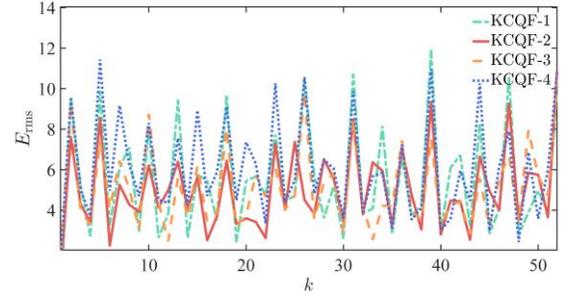

Fig. 1. Performance comparison for different numbers of KCQs.

Next, we will compare the KCQF with some existing filters. This example calculates the $\overline{E_{rms}}$ of five filters, namely PGM-UT (particle Gaussian mixture with unscented transform), PGM (particle Gaussian mixture) [37], PF-RR (residual resampling) [39], PF-SR (stratified resampling) [40], and UKF filters, and compare these with the $\overline{E_{rms}}$ calculated by the KCQF-2 and KCQF-3 methods. The results are plotted in Fig. 2. To ensure fairness, the number of samples the number of particles $N_s$ is uniformly set to 50. At the same time, this example also compares the CPU times taken by different filters during operation. The results are plotted in Fig. 3.

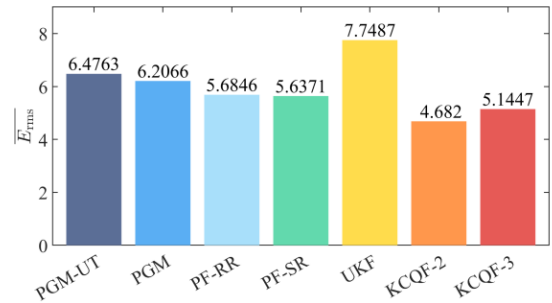

Fig. 2. $\overline{E_{rms}}$ comparison for different filters.

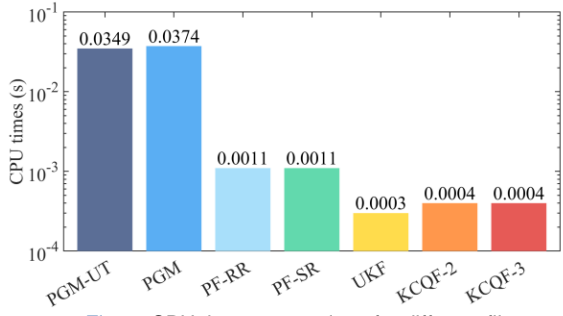

Fig. 3. CPU times comparison for different filters.

The results shown in Fig. 2 indicate that the proposed KCQF-2 and KCQF-3 filters have good estimation accuracy. The $\overline{E}_{\text{rms}}$ of KCQF-2 and KCQF-3 are better than that of other comparative filters. Furthermore, it is not difficult to see from Fig. 2 that for this nonlinear model, KCQF-2 yields the best calculation result. Figure 3 clearly demonstrates that the KCQF-2 and KCQF-3 filters has excellent computational efficiency, with CPU times only slightly lower than the UKF filter, and significantly less than other compared filters. We further plot the convergence curves of KCQF-2 and KCQF-3 with different numbers of samples in Fig. 4. It is not difficult to see from Fig. 4 that when the number of samples $N_s = 50$, the $\overline{E}_{\text{rms}}$ calculated by KCQF-2 is already less than 5. As the number of samples gradually increases, the $\overline{E}_{\text{rms}}$ obtained from KCQF-2 and KCQF-3 gradually decrease in an oscillatory manner and converge to $\overline{E}_{\text{rms}} = 4.5$.

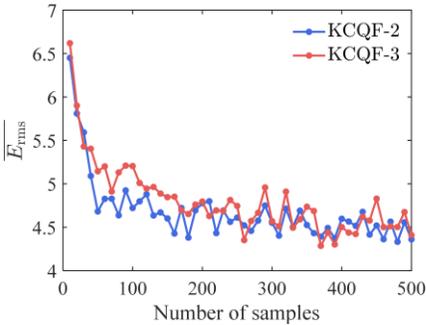

Fig. 4. Convergence curves of KCQF-2 and KCQF-3.

### B. Non-Gaussian, non-Markovian nonlinear example

The second example continues to employ the state and observation models described by (32) in Subsection IV.A, and retains the initial conditions and observation noise used in that subsection. However, the difference lies in the process noise, which is no longer Gaussian white noise but is assumed to be a non-Gaussian with a mean of $\overline{w} = 0$ and a variance of 10. Utilizing the Karhunen-Loeve (K-L) expansion [41] to establish the process noise, $w_{k+1}$ can be expressed as:

$$w_{k+1} = \overline{w} + \sum_{n=1}^{M} \xi_n \sqrt{\lambda_n} f_{n,k+1}, \quad (35)$$

where, $\xi_n \in \left[-\sqrt{30}, \sqrt{30}\right]$ is a uniform random variable, $M = 6$ is the expansion order of the K-L expansion. $f_{n,k+1}$ and $\lambda_n$ are the $n$-th eigenfunctions and eigenvalues of the autocorrelation function $\rho(i,j)$, satisfying:

$$\rho(i,j)f_{n,j} = \lambda_n f_{n,j}, 1 \le i,j,n \le K, \lambda_1 > \lambda_2 > \cdots > \lambda_6$$
$$\sum_{i=1}^{K} f_{m,i} f_{n,i} = \delta_{m,n}, \sum_{i=1}^{K} f_{m,i} \rho(i,j) f_{n,i} = \lambda_n \delta_{m,n} \quad , (36)$$

in which $\delta_{m,n}$ is the Kronecker delta function. In this example, the autocovariance function $\rho(i,j)$ is defined as:

$$\rho(i,j) = \exp\left[-\left(\frac{i-j}{15}\right)^2\right]. \quad (37)$$

To illustrate the non-Gaussian and non-Markovian characteristics of the problem, we conducted random sampling on $\xi$ 1000000 times, based on the results, plotted the PDF of $w_{26}$, as depicted by the red solid line in Fig. 5(a). The blue dashed line in Fig. 5(a) represents the Gaussian PDF plotted based on the mean and variance of $w_{26}$. It is evident from Fig. 5(a) that $w_{26}$ does not follow a Gaussian distribution. To further illustrate that the problem is not a Markov process, we have used the same method to plot the conditional PDF $p_{x_2|x_1}(x_2|x_1)$ and $p_{x_2|x_1,x_0}(x_2|x_1,x_0)$ under conditions $x_0 = -0.5$ and $x_1 = -0.2$, respectively. As shown in Fig. 5(b). It is evident from Fig. 5(b) that $p_{x_2|x_1}(x_2|x_1) \ne p_{x_2|x_1,x_0}(x_2|x_1,x_0)$, indicating that the state $x_k$ is a non-Markovian random process.

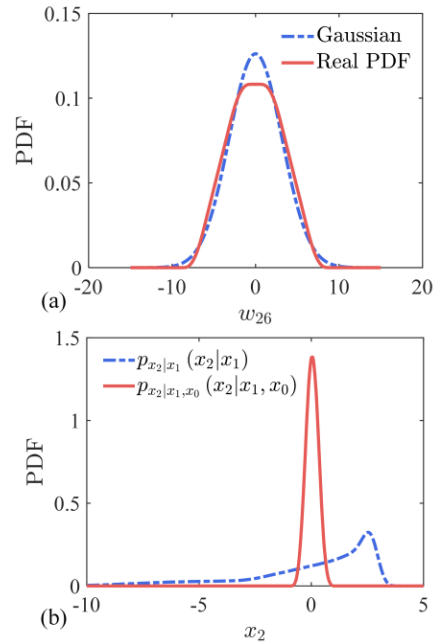

Fig. 5. Non-Gaussian and non-Markovian characteristics of the problem: (a) PDF curve of $w_{26}$; (b) Conditional PDFs $p_{x_2|x_1}(x_2|x_1)$ and $p_{x_2|x_1,x_0}(x_2|x_1,x_0)$ under conditions $x_0 = -0.5$ and $x_1 = -0.2$.

In this example, the proposed KCQF is used to analyze the nonlinear problem of the non-Gaussian, non-Markovian process. Different number of key conditions $d$ are selected to compute the $\overline{E_{\text{rms}}}$, with 50 MC runs still being conducted, and the results are listed in Table 1. From Table 1, it can be observed that different $d$ have a certain impact on the accuracy of KCQF. As the $d$ increases, $\overline{E_{\text{rms}}}$ shows a trend of decreasing first and then increasing, and the $\overline{E_{\text{rms}}}$ of KCQF-3 is the best among all filters. When $d=2$ and 3, the $E_{\text{rms}}$ is superior to that when $d=1$, this indicates that considering more key measurements leads to better estimation accuracy than considering only one key measurement. When $d \geq 4$, $\overline{E_{\text{rms}}}$ starts to increase slightly. This phenomenon corroborates again the point made in section III.B: If too many measurement values are considered, the $p_{\beta_{k+1}}(\beta_{k+1})$ will become too small, leading to both the numerator and the denominator in (18) and (19) to become small, thereby increasing the error in numerical computation.

Currently, some outstanding filters, such as the EKF, UKF, CKF, and the PF, are widely utilized. EKF, UKF, and CKF typically assume that the process noise is Gaussian; the PF generally presume that the system states follow a Markov process. However, the problem addressed in this example involves a random process that is neither Gaussian nor Markovian, and neglecting these characteristics will lead to a significant decrease in prediction accuracy. To illustrate this point, this example employs seven types of filters, EKF, UKF, CKF, PF-LVR, PF-STR, PGM-UT, and PGM, as comparative filters for computation. The results of these different filters are also presented in Table 1. From Table 1, it can be observed that due to the neglect of the non-Gaussian and non-Markovian characteristics of the problem by EKF, UKF, and CKF, their estimation results are inaccurate, with very large $\overline{E_{\text{rms}}}$ values. Among the three methods, UKF has the best accuracy, but its $\overline{E_{\text{rms}}}$ is still significantly larger than the $\overline{E_{\text{rms}}}$ of the proposed KCQF. PF-LVR, PF-STR, PGM-UT and PGM can take into account the non-Gaussian characteristic of the problem. However, due to the neglect of the non-Markovian characteristics, the calculated $\overline{E_{\text{rms}}}$ for these methods is much larger than that of the proposed KCQF. We also compared the computation times of different filters, as shown in Table 1, where the computation times are the averages obtained from performing 50 MC runs. It can be observed that the computation times of different filters do not differ significantly. Summarizing from Fig. 6 and Table 1, the proposed KCQF can take into account the non-Gaussian and non-Markovian characteristics of the filtering problem, achieving more accurate state estimation.

TABLE I
COMPARISON OF $\overline{E_{\text{rms}}}$ AND CPU TIMES CALCULATED BY DIFFERENT FILTERS

| Filter | KCQF-1 | KCQF-2 | KCQF-3 | KCQF-4 | KCQF-5 | KCQF-6 | KCQF-7 |
|---|---|---|---|---|---|---|---|
| $\overline{E_{\text{rms}}}$ | 4.7965 | 2.1213 | 1.8797 | 2.2884 | 2.3061 | 2.4685 | 2.5653 |
| CPU times (s) | 8.81e-4 | 8.92e-4 | 8.64e-4 | 8.87-4 | 8.74e-4 | 9.70e-4 | 0.0011 |

| Filter | PF-RR | PF-SR | PGM-UT | PGM | EKF | UKF | CKF |
|---|---|---|---|---|---|---|---|
| $\overline{E_{\text{rms}}}$ | 5.1707 | 5.5060 | 6.0641 | 6.2181 | 22.2849 | 8.7809 | 22.8300 |
| CPU times (s) | 0.0015 | 0.0015 | 0.0380 | 0.0349 | 9.50e-4 | 8.08e-4 | 0.0011 |

We further plot the convergence curves of KCQF-2, KCQF-3 and KCQF-4 with different numbers of samples in Fig. 6. It is not difficult to see from Fig. 6 that when the number of samples $N_s = 50$, the $\overline{E_{\text{rms}}}$ calculated by KCQF-3 is already less than 2. As the number of samples gradually increases, the $\overline{E_{\text{rms}}}$ obtained from KCQF-2, KCQF-3 and KCQF-4 gradually decrease in an oscillatory manner and converge to $\overline{E_{\text{rms}}} = 1.75$.

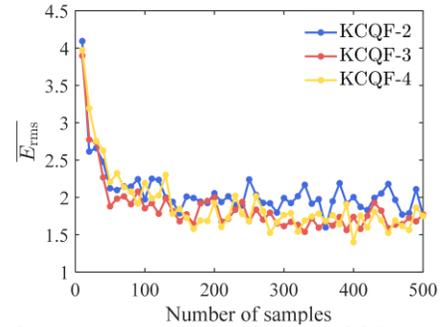

Fig. 6. Convergence curves of KCQF-2, KCQF-3 and KCQF-4.

## V. CONCLUSION

This paper studies the estimation of state given measurement conditions. We first theoretically observe, without involving approximations such as Gaussian distributions or Markov processes, that when considering all measurement conditions, the numerator and denominator of the state estimation quotient-form expression tend to zero simultaneously over time, making numerical calculations unstable. This observation motivates us to propose the idea of estimating the state based on key measurement conditions, rather than all measurement conditions. According to this idea, using the principle of probability conservation, we have derived the corresponding integral quotient-form expressions for the conditional PDF, mean, and variance of states based on key measurement conditions, and employed the MC method to calculate these expressions, thereby constructing a novel key conditional quotient filter (KCQF). KCQF uses key conditions to estimate states, avoiding the numerical difficulty that the numerator and denominator tend to zero as time increases. Two nonlinear numerical examples were given to demonstrate the superior estimation performance of KCQF, compared to other filters. In the future, we plan to: 1) Extend the concept of key measurement conditions to other filters that require sampling, such as particle filter; 2) Extend our research to address filtering problems involving uncertain noises whose PDF is unknown and only its interval bounds are known.

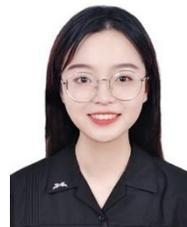

**Yuelin Zhao** is working as a postgraduate student in Computational mechanics with the faculty of vehicle engineering and mechanics, Dalian University of Technology. Her research interests include the construction and application of the low-discrepancy samples, filtering algorithms and uncertainty quantification.

1111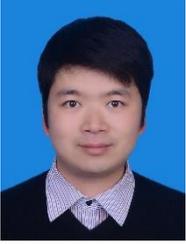 **Feng Wu** received his B.Sc degrees in Civil Engineering from Guangxi University in 2007. And from the same university, he received the M.Eng in Structural Engineering in 2010. He obtained the Ph.D degree in Engineering Mechanics from Dalian University of Technology in 2015. From 2015 to 2017, he worked as a post-doctor in the School of Naval Architecture & Ocean Engineering at Dalian University of Technology. From 2018-2023, he served as an associate professor in the Faculty of Vehicle Engineering and Mechanics at Dalian University of Technology. Since January 2024, he has been a professor in the School of Mechanics and Aerospace Engineering at Dalian University of Technology. His research direction is the dynamics and control, filtering algorithms and uncertainty quantification. He is an editorial board member of Computational Mechanics and Chinese Journal of Applied Mechanics.

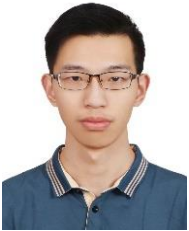 **Li Zhu** is working as a postgraduate student in Computational mechanics with the faculty of vehicle engineering and mechanics, Dalian University of Technology. His main research interests are filtering algorithms and uncertainty quantification.